\documentstyle[12pt,elsart12]{article}
\newcommand{\be}{\begin{equation}}
\newcommand{\ee}{\end{equation}}
\begin{document}
\pagestyle{empty}
\centerline{\bf{INVERSE COMPTON SCATTERING}}
\centerline{\bf{ON LASER BEAM AND MONOCHROMATIC}}
\centerline{\bf{ISOTROPIC RADIATION}}
\vskip 0.5cm
\centerline{\bf{ Daniele Fargion,$^{1,2}$ Rostislav V. Konoplich$^{1,2}$ and 
Andrea Salis$^1$ }}
\centerline{}
\centerline{$^1$ {\it Dipartimento di Fisica,
Universita'~di~Roma~"La~Sapienza", }}
\centerline{\it P.le~A.~Moro~2,~00185~Rome,~Italy. }
\centerline{$^2$ {\it INFN-Sezione~di~Roma~I,~c/
o~Dipartimento~di~Fisica,}}
\centerline{\it~Universita'~di~Roma~"La~Sapienza",~00185~Rome,~Italy}
\centerline{}
\begin{abstract}
Most of the known literature on Inverse Compton Scattering (ICS) is 
based on earliest theoretical attempts and later approximations led by 
F.C.Jones and J.B.Blumenthal. We found 
an independent and more general analytical procedure which provide both 
$\it{relativistic}$ and $\it{ultrarelativistic}$ limits for ICS. These new 
analytical expressions can be derived in a straightforward way and they 
contain the previously reminded Jones' results. Our detailed solutions may be 
probed by already existing as well future ICS experiments. 
\end{abstract}
\vskip 1cm
\begin{flushleft}
e-mail:FARGION@ROMA1.INFN.IT\\
e-mail:KONOPLIC@ROSTI.MEPHI.MSK.SU\\
e-mail:SALIS@ROMA1.INFN.IT\\
Tel:39 6 4991-4287 (Fargion)\\
fax:4957697
\end{flushleft}
\newpage
\pagestyle{plain}
\setcounter{page}{1}
\section*{Introduction}
\label{sec:intro}

The ICS plays a relevant role in a variety of recent high energy astrophysics 
(cosmic ray lifetime, gamma astronomy, gamma jets, ultrahigh energy cosmic 
rays (UHECR),...) as well as in high energy 
physics at LEP I, LEP II and linear accelerators .
In particular the ICS of cosmic rays onto electromagnetic fields, either
cosmological Black Body Radiation (BBR) at $T\approx 2.73~K$ or diffused gray 
body as interstellar lights and radio waves or even nearly stationary 
magnetic fields, is source of high energy photons (X, gamma ,...) which we may
observe as diffused or smeared pointlike gamma sources. The ICS onto BBR 
plays an important role in the complex connection between cosmic rays and 
gamma spectra and in the gamma burst puzzle. In this paper we derive the 
general ICS formulae for the interaction between a monochromatic electron beam 
and a monochromatic photon beam, as a laser, either unidirectional or 
isotropic. We shall show that our results generalize and slightly correct 
previous known ones [1-5]. The complexity of the ICS onto a Planckian spectrum 
has already been experimentally verified [6,7] as well as successfully 
theoretically predicted [8,9].

\section*{The ICS on a monochromatic and unidirectional photon beam}

To find the ICS spectrum we consider first the photon beam distribution in 
the Laboratory Frame (LF) (not to be confused with the electron rest frame EF) 
where this beam distribution is unidirectional and 
monochromatic; then we transform it in the (EF) where it still 
is unidirectional and monochromatic; in that frame (EF) we consider the normal 
Compton 
or Thomson scattering and finally we transform back the resulting diffused 
differential photon number to the LF.\par
Let us consider such a monochromatic and unidirectional photon beam in the LF
: the differential number density per unit energy $\epsilon_o$ and solid angle 
$\Omega_o$ can be written as:  
\be
\frac{dn_o}{d\epsilon_o d\Omega_o}=n_o\delta(\epsilon_o-
\hat{\epsilon}_o)\delta(\cos\theta_o-\cos\hat{\theta}_o)\delta(\varphi_o-
\hat{\varphi}_o)
\ee
where $\hat{\theta}_o$ is the incident 
angle between the electron beam and photon beam 
directions and $\hat{\epsilon}_o$ is the initial photon energy in the LF. Our 
sources have an idealized beam with no spread both in angle and in energy. We 
transform now this distribution to the EF by standard Lorentz relations 
choosing as z axis the direction coincident with the electron momentum and 
reminding that $dn_o/d\epsilon_o$ is a relativistic invariant [2]. In the 
following we label by a $^*$ the quantities related to the electron frame EF, 
so we have
\be
\cos\theta_o=\frac{\cos\theta_o^*+\beta}{1+\beta\cos\theta_o^*
}~,~\varphi_o=\varphi_o^*~,~
\epsilon_o=\gamma\epsilon_o^*(1+\beta\cos\theta_o^*)~~~~.
\ee
where $\beta$ is the adimensional electron velocity and $\gamma$ is the 
corresponding Lorentz factor. 
Using $\delta$ function properties the number density distribution, 
in the EF, becomes
$$
\frac{dn_o^*}{d\epsilon_o^* d\Omega_o^*}=n_o \gamma 
(1-\beta\cos\hat{\theta}_o)\cdot
$$
\be
\cdot\delta[\epsilon_o^* -
\gamma\hat{\epsilon}_o (1-\beta\cos\hat{\theta}_o)]\delta[\cos\theta_o^* -
\hat{C}_o]\delta(\varphi_o^*-\hat{\varphi}_o)
\ee
where $\hat{C}_o=\theta_o^*=(\cos\hat{\theta}_o-\beta)/(1-
\beta\cos\hat{\theta}_o)$ is the function describing the boosted cosine angle 
$\theta_o^*$. Now we have to scatter and diffuse the photons in the EF; the 
function describing the differential number of diffused photons can be 
obtained as follows: 
\begin{equation}
\frac{dN_1^*}{dt_1^*d\epsilon_1^*d\Omega_1^*d\epsilon_o^*d\Omega_o^*}=\frac
{dn_o^*}{d\epsilon_o^*d\Omega_o^*}\frac{d\sigma_C}{d\epsilon_1^*d\Omega_1^*}c
\end{equation}
where $\frac{d\sigma_C}{d\epsilon_1^*d\Omega_1^*}$ is the Klein-Nishina 
differential cross section 
$$
\frac{d\sigma_C}{d\epsilon_1^*d\Omega_1^*}=\frac{r_o^2}{2}\Big(
\frac{\epsilon_1^*}{\epsilon_o^*}\Big)^2\Big(\frac{\epsilon_1^*}{\epsilon_o^*}
+\frac{\epsilon_o^*}{\epsilon_1^*}-\sin^2\theta_{sc}^*\Big)\delta\Big(
\epsilon_1^*-\frac{\epsilon_o^*}{1+\epsilon_o^*(1-\cos\theta_{sc}^*)/mc^2}\Big
)~~.
$$
For most of the real ICS processes in present laboratories energetics it is 
possible to approximate the Klein-Nishina cross section by the Thomson 
cross section, $\it{i.e.}$ $\epsilon_o^*\ll mc^2$, so we can consider 
\be
\frac{d\sigma_T}{d\epsilon_1^*d\Omega_1^*}=\frac{r_o^2}{2} (1+\cos^2\theta_{sc}
^*)\delta(\epsilon_1^* -\epsilon_o^*)
\ee
where $c$ is the speed of light. The dependence of $\sigma_T$ by the inverse 
square 
electron mass leads us to consider mainly electron bunches. However our 
results can 
be applied also to protons where we have just a suppression factor $(m_e/
m_p)^2$. The scattering angle in the EF must be expressed as a function of the 
other angles involved, $\it{i.e.}$ the incoming $\theta_o^*$,$\varphi_o^*$ and 
the outcoming $\theta_1^*$,$\varphi_1^*$ angles
\be
\cos\theta_{sc}^*=\sin\theta_o^*\sin\theta_1^*(\cos\varphi_o^*\cos\varphi_1^*+
\sin\varphi_o^*\sin\varphi_1^*)+\cos\theta_o^*\cos\theta_1^*~~~~.
\ee 
It is interesting to notice that the distribution number 
$\frac{dN_1}{dt_1^*d\epsilon_1^*d\Omega_1^*}$ cannot be associated to any 
effective number density because of its intrinsic inhomogeneous nature. 
Finally we obtain the differential photon number per unit energy and solid 
angle in the LF by the 
inverse Lorentz transformations and we write it in the following integral form:
\be
\frac{dN_1}{dt_1d\epsilon_1d\Omega_1}=\int_{\Omega_o^*}\int_{\epsilon_o^*}
\frac{dN_1^*}{dt_1^*d\epsilon_1^*d\Omega_1^*d\epsilon_o^*d\Omega_o^*
}\frac{dt_1^*}{dt_1}\frac{d\epsilon_1^*}{d\epsilon_1}\frac{d\Omega_1^*
}{d\Omega_1}d\Omega_o^*d\epsilon_o^*~~~~~~.
\ee
By sostitution of previous eqs.(4,6) and since 
$\frac{dt_1^*}{dt_1}\frac{d\epsilon_1^*}{d\epsilon_1}\frac{d\Omega_1^*
}{d\Omega_1}=\frac{1}{\gamma^2(1-\beta\cos\theta_1)}$ the integral can be cast 
into the form
$$
\frac{dN_1}{dt_1 d\epsilon_1d\Omega_1}=\frac{n_o r_o^2 c}{2\gamma}\frac{(1-
\beta\cos\hat{\theta}_o)}{(1-\beta\cos\theta_1)}\int_{\epsilon_o^*}
\int_{\Omega_o^*}(1+\cos^2\theta_{sc}^*)\delta(\epsilon_1^* -\epsilon_o^*)
\cdot
$$
\be
\cdot\delta[\epsilon_o^* -\gamma\hat{\epsilon}_o (1-\beta\cos\hat{\theta}_o)]
\delta(\cos\theta_o^*-\hat{C}_o)\delta(\varphi_o^*-\hat{\varphi}_o) 
d\epsilon_o^* d\Omega_o^*~~~~.
\ee
We perform the integrals and the final form for the differential photon number 
per unit energy and solid angle in the Laboratory Frame becomes
$$
\frac{dN_1}{dt_1d\epsilon_1d\Omega_1}=\frac{n_o r_o^2 
c}{2\beta\gamma^2\epsilon_1}\frac{(1-\beta\cos\hat{\theta}_o)}{(1-
\beta\cos\theta_1)}\Bigg(1+C^2_1 \hat{C}^2_o+(1-C^2_1)(1-\hat{C}^2_o)(
\cos\varphi_1\cos\hat{\varphi}_o+
$$
$$
+\sin\varphi_1\sin\hat{\varphi}_o)^2
+2C_1 \hat{C}_o (1-C^2_1)^{1/2} (1-\hat{C}^2_o)^{1/2} 
(\cos\varphi_1\cos\hat{\varphi}_o+
\sin\varphi_1\sin\hat{\varphi}_o)\Bigg)\cdot
$$
\be
\cdot\delta\Big[\cos\theta_1-\frac{1}{\beta}\Big(1-
\frac{\hat{\epsilon}_o}{\epsilon_1}(1-\beta\cos\hat{\theta}_o)\Big)
\Big]~~.
\ee
where $C_1=\cos\theta_1^*=(\cos\theta_1-\beta)/(1-\beta\cos\theta_1)$.
The final ICS spectrum can be obtained by integrating over $\Omega_1$ the 
previous equation and the result is:
\be
\frac{dN_1}{dt_1 d\epsilon_1}=\frac{\pi n_o r_o^2 
c}{2\beta\gamma^2\hat{\epsilon}_o}\Big[3-\hat{C}_o^2+(3\hat{C}_o^2-1)
\frac{1}{\beta^2}\Big(\frac{\epsilon_1}
{\gamma^2\hat{\epsilon}_o(1-\beta\cos\hat{\theta}_o)}-1\Big)^2\Big].
\ee
Let us notice that this energy distribution does not depend on the initial 
azimuthal angle $\hat{\varphi}_o$ due to the axial symmetry of the problem. 
The $\epsilon_1$ dependence shows that the original monochromatic and 
unidirectional photon spectrum has been spread into a final parabolic 
function. We show below its behaviour for some arbitrary parameters (fig.1-5). 
The spectrum is bound in energy by relativistic kinematics arguments and its 
extreme $\epsilon_1$ allowed values are 
$\epsilon_{1min}=\hat{\epsilon}_o(1-\beta\cos\hat{\theta}_o)/(1+\beta)$ and 
$\epsilon_{1max}=\hat{\epsilon}_o(1-\beta\cos\hat{\theta}_o)/(1-\beta)$. The 
$\theta_1$ angle is simply varying into the range $[0,\pi]$ and, as usual, the 
most populate angular region of the high energy ICS beam is contained inside a 
thin cone whose aperture is of order $1/\gamma$. If we substitute in eq.(10) 
$\epsilon_{1min}$ and $\epsilon_{1max}$ we get an equal height for the 
parabolic spectrum extremes:
\be
\frac{dN_1}{dt_1 d\epsilon_1}_{\mid \epsilon_{1(min,max)}}=\frac{\pi n_o r_o^2 
c}{\beta\gamma^2\hat{\epsilon}_o}[1+\hat{C}_o^2]~~~~~.
\ee
The spectrum has a minimum for $\epsilon_1=\gamma^2\hat{\epsilon}_o(1-
\beta\cos\hat{\theta}_o)$ and in this point its value is:
\be
\frac{dN_1}{dt_1 d\epsilon_1}_{\mid min}=\frac{\pi n_o r_o^2 
c}{2\beta\gamma^2\hat{\epsilon}_o}[3-\hat{C}_o^2]~~~~.   
\ee 
The above expressions are the rigorous analytical spectra for Thomson ICS 
onto a 
monochromatic and unidirectional photon beam. From eq.(9) we can also get 
the angular distribution of scattered photons and the total rate number:
$$
\frac{dN_1}{dt_1d\Omega_1}=\frac{n_o r_o^2 c}{2\gamma^2}\frac{(1-
\beta\cos\hat{\theta}_o)}{(1-\beta\cos\theta_1)^2}
[1+C^2_1 \hat{C}_o^2+(1-C_1^2)(1-\hat{C}_o^2)(\cos\varphi_1\cos\hat{\varphi}_o+
$$
\be
+\sin\varphi_1\sin\hat{\varphi}_o)^2+
2C_1 \hat{C}_o (1-C_1^2)^{1/2} (1-\hat{C}_o^2)^{1/2} 
(\cos\varphi_1\cos\hat{\varphi}_o+\sin\varphi_1\sin\hat{\varphi}_o)]
\ee
and
\be
\frac{dN_1}{dt_1}=\sigma_Tn_oc(1-\beta\cos\hat{\theta}_o)~~~~~~~~.
\ee
The spectrum in eq.(10) may be tested on known records of ICS. Our eq.(10) also 
contains the non relativistic limit where, for $\beta\rightarrow 0$, one must 
recover the correct initial Dirac delta function 
$\delta(\epsilon_1-\hat{\epsilon}_o)$.

\section*{ICS on a monochromatic and unidirectional electron beam beyond 
the Thomson limit}

Let us consider now the ICS in the framework of quantum electrodynamics in 
order to obtain the exact Compton result for the corresponding particle 
distribution in the case of a monochromatic and unidirectional photon beam. 
Considering two Feynman diagrams which contribute to this process the 
standard calculations give us the following expressions for the matrix element 
(in this section we use $\hbar=c=1$):
\be
\mid\bar{M}\mid^2= 2^5 \pi^2 r_o^2 m^2\Bigg(\Big[m^2\Big(\frac{1}{\kappa_o p_o}
-\frac{1}{\kappa_1 p_o}\Big)+1\Big]^2+\frac{\kappa_1 p_o}{\kappa_o p_o}+
\frac{\kappa_o p_o}{\kappa_1 p_o}-1\Bigg)
\ee
where $\kappa_o p_o=\hat{\epsilon}_o m\gamma(1-\beta\cos\hat{\theta}_o)$, 
$\kappa_1 p_o=\epsilon_1 m\gamma(1-\beta\cos\theta_1)$. The corresponding 
cross section is given by
\be
d\sigma=\frac{1}{2^6\pi^2}\frac{\mid\bar{M}\mid^2}{(p_o\kappa_o)}\delta^{(4)}
(p_o+\kappa_o-p_1-\kappa_1)\frac{d^3\kappa_1}{\epsilon_1}\frac{d^3 
p_1}{E_1}~~~~~~~.
\ee
In the case of colliding beams the number of collisions per second can be 
obtained from the following relation
\be
d\dot{N}_1=Ld\sigma
\ee
where L is the luminosity which is defined by [10]
\be
L=n_o n_1 V(1-\beta\cos\hat{\theta}_o)
\ee
V is the unit volume in the LF. In our case $n_1 V=1$, the density $n_o$ was 
defined in eq.(1). Now integrating eq.16 over the corresponding variables we 
obtain the following exact expression for the angular distribution of 
scattered photons:
$$
\frac{dN_1}{dt_1 d\Omega_1}=n_o\frac{r_o^2}{2}\frac{m^2\epsilon_1^2}{(
p_o\kappa_o)^2}(1-\beta\cos\hat{\theta}_o)\cdot
$$
\be
\cdot\Bigg(\Big[m^2\Big(\frac{1}{\kappa_o p_o}-\frac{1}{\kappa_1 
p_o}\Big)+1\Big]^2+\frac{\kappa_1 p_o}{\kappa_o p_o}+\frac{\kappa_o 
p_o}{\kappa_1 p_o}-1\Bigg)
\ee
where 
$$
\epsilon_1=\frac{\hat{\epsilon}_o m\gamma(1-\beta\cos\hat{\theta}_o)}
{\gamma m(1-\beta\cos\theta_1)+\hat{\epsilon}_o(1-\cos\theta_{sc})}
$$
and the scattering angle is defined as in eq.(6) but in the LF. The energy 
spectrum of ICS is given by
$$
\frac{dN_1}{dt_1 d\epsilon_1}=n_o\pi r_o^2\frac{m}{\hat{\epsilon}_o\gamma}
\Bigg(\frac{\gamma m^5}{\hat{\epsilon}_o^2 A^{3/
2}}\Big[\epsilon_1(\hat{\epsilon}_o+m\gamma\beta\cos\hat{\theta}_o)+
$$
$$
+\beta(
\hat{\epsilon}_o\cos\hat{\theta}_o+m\gamma\beta)(m\gamma(1-
\beta\cos\hat{\theta}_o)-\epsilon_1)\Big]+\frac{\kappa_o p_o}{\hat{\epsilon}_o 
A^{1/2}}\Big[1-\frac{2 m^2}{\kappa_o p_o}-2\Big(\frac{m^2}{\kappa_o p_o}\Big)
^2\Big]+
$$
$$
+\frac{1}{B^{1/2}}\Big[\Big(\frac{m^2}{\kappa_o p_o}\Big)^2+\frac{2 
m^2}{\kappa_o p_o}+\frac{\epsilon_1\gamma m}{\kappa_o p_o}\Big]+
\frac{m\gamma\beta}{(\kappa_o p_o)B^{3/2}}(\hat{\epsilon}_o\cos\hat{\theta}_o+
m\gamma\beta)\cdot
$$
\be
\cdot(\kappa_o p_o-\epsilon_1\gamma m-\epsilon_1\hat{\epsilon}_o)\Bigg
)
\ee
where
$$
A=m^2\Bigg(\gamma^2\Big[m\gamma\beta(1-\beta\cos\hat{\theta}_o)+\epsilon_1(
\cos\hat{\theta}_o-\beta)\Big]^2+\epsilon_1^2(1-\cos^2\hat{\theta}_o)\Bigg)
$$
$$
B=\hat{\epsilon}_o^2+m^2\gamma^2\beta^2+2\hat{\epsilon}_o 
m\gamma\beta\cos\hat{\theta}_o
$$
and
$$
\frac{\hat{\epsilon}_o m\gamma(1-\beta\cos\hat{\theta}_o)}
{m\gamma+\hat{\epsilon}_o+\sqrt{B}}\leq\epsilon_1\leq\frac{\hat{\epsilon}_o 
m\gamma(1-\beta\cos\hat{\theta}_o)}{m\gamma+\hat{\epsilon}_o-\sqrt{B}}
$$
These exact expressions reproduce and confirm all the results of previous 
section as Thomson limit cases. Therefore eq.(10) can be used for further 
calculations of ICS on an isotropic monoenergetic background within the 
Thomson limit. Expression (20) may find application when considering the most 
energetic accelerators where the colliding electron and photon beams are of 
high energy. The behaviour of the analytical formula eq.(20) is shown in 
fig.6. In comparison with the spectrum discussed above we note that the 
Thomson "parabolic" behaviour becomes in the present Compton case an 
asymmetric curve which (in extreme ultrarelativistic regime where 
$\hat{\epsilon}_o\gamma\gg m$) leads asymptotically to a nearly peaked curve 
at energy $\epsilon_1\simeq m\gamma$.

\section*{The ICS onto a monochromatic and isotropic photon spectrum}

In ref.[2] F.C. Jones found the ultrarelativistic ICS spectrum resulting 
from the 
interaction between high energy electrons and an isotropic and monochromatic 
photon spectrum. We show that his result can be obtained and slightly 
corrected by integrating eq.(10) over all permitted $\hat{\theta}_o$ angles. 
Our analytical and exact result can be set in a very compact form and it is 
derived from the following integral 
$$
\frac{dN_{1is}}{dt_1 d\epsilon_1}=
\frac{1}{4\pi}\int_{\Omega_o}\frac{dN_1}{dt_1 d\epsilon_1} d\Omega_o=
$$
\be
=\frac{\pi n_o r_o^2 c}{4\beta\gamma^2\hat{\epsilon}_o}
\int_{\hat{\theta}_{min}}^{\hat{\theta}_{max}} 
\Big[3-\hat{C}_o^2+(3\hat{C}_o^2-1)\frac{1}{\beta^2}\Big(
\frac{\epsilon_1}{\gamma^2\hat{\epsilon}_o(1-\beta\cos\hat{\theta}_o)}-
1\Big)^2\Bigg] \sin\hat{\theta}_o d\hat{\theta}_o
\ee
The upper and lower limits of this integral can be found from the inequality 
$\epsilon_{1min}\leq\epsilon_1\leq\epsilon_{1max}$ where the minimum and 
maximum energies of scattered photons were defined in the first section. Thus 
we obtain
$$
\frac{1-\beta}{1+
\beta}\hat{\epsilon}_o\leq\epsilon_1\leq\hat{\epsilon}_o~~~~~~~\frac{1}{\beta}
\Big[1-\frac{\epsilon_1}{\hat{\epsilon}_o}(1+\beta)
\Big]\leq\cos\hat{\theta}_o\leq 1
$$
$$
\hat{\epsilon}_o\leq\epsilon_1\leq\frac{1+\beta}{1-
\beta}\hat{\epsilon}_o~~~~~~~-1\leq\cos\hat{\theta}_o\leq\frac{1}{\beta}\Big[1
-\frac{\epsilon_1}{\hat{\epsilon}_o}(1-\beta)\Big]~~.
$$
As one can see from these expressions only photons incoming within a thin cone 
in the direction of the electron beam contribute to the lowest energy final 
photons meanwhile photons moving in the opposite direction contribute to the 
highest energy part of the spectrum. An elementary calculation gives
$$
\frac{dN_{1is}}{dt_1 d\epsilon_1}=\frac{\pi n_o r_o^2 
c}{4\beta^4\gamma^2\hat{\epsilon}_o}\Bigg(\Big[3\beta^2-2+
\frac{3}{\beta^2}\Big]x+\frac{2}{\gamma^2}\Big[\Big(1-\frac{3}{\beta^2}\Big)
\Big(1+\frac{\epsilon_1}{\hat{\epsilon}_o}\Big)\Big]\ln x+
$$
$$
+\frac{1}{\gamma^4}\Big[\Big(1-\frac{3}{\beta^2}\Big)\Big(1+
\frac{\epsilon_1^2}{\hat{\epsilon}_o^2}\Big)-
\frac{12\epsilon_1}{\beta^2\hat{\epsilon}_o}\Big]\frac{1}{x}+
$$
\be
+\frac{3\epsilon_1}{\beta^2\gamma^6\hat{\epsilon}_o}\Big(1+\frac{\epsilon_1}
{\hat{\epsilon}_o}\Big)\frac{1}{x^2}
-\frac{\epsilon_1^2}{\beta^2\gamma^8\hat{\epsilon}_o^2}\frac{1}{x^3}\Bigg)
_{x_{min}}^{x_{max}}
\ee
where $x=(1-\beta\cos\hat{\theta}_o)$. We have now two different regions 
depending on the value assumed by the ratio $\epsilon_1/\hat{\epsilon_o}$: if 
$\frac{(1-\beta)\hat{\epsilon}_o}{1+\beta}\leq\epsilon_1\leq\hat{\epsilon}_o$ 
then 
$x_{max}=[\epsilon_1 (1+\beta)/\hat{\epsilon}_o]$ and $x_{min}=(1-\beta)$; if 
$\hat{\epsilon}_o\leq\epsilon_1\leq\frac{(1+\beta)\hat{\epsilon}_o}{1-\beta}$ 
then 
$x_{max}=(1+\beta)$, $x_{min}=[\epsilon_1 (1-\beta)/\hat{\epsilon}_o]$. The two 
separate formulae, left and right, must vanish respectively at $\epsilon_1=
\hat{\epsilon}_o(1-\beta)/(1+\beta)$ and $\epsilon_1=\hat{\epsilon}_o(1+\beta)
/(1-\beta)$ and they must obviously coincide for $\epsilon_1=\hat{\epsilon}_o$.
Labelling by L the left hand side and by R the right hand 
side, with respect to the value $\epsilon_1=\hat{\epsilon}_o$, the two 
formulae, exact in the whole range of allowable values for the ratio 
$\epsilon_1/\hat{\epsilon}_o$, become
$$
\frac{dN_{1is~L}}{dt_1d\epsilon_1}=\frac{\pi n_o r_o^2 
c}{4\beta^6\gamma^2\hat{\epsilon}_o}\Bigg(\frac{\epsilon_1}{\hat{\epsilon}_o}(
1+\beta)\Big[\beta(\beta^2+3)+\frac{1}{\gamma^2}(9-4\beta^2)\Big]+
$$
$$
+(1-\beta)\Big[\beta(\beta^2+3)-\frac{1}{\gamma^2}(9-4\beta^2)\Big]+
$$
$$
-\frac{2}{\gamma^2}(3-\beta^2)\Big(1+\frac{\epsilon_1}{\hat{\epsilon}_o}\Big)
\ln\Big[\frac{\epsilon_1(1+\beta)}{\hat{\epsilon}_o(1-\beta)}\Big]-
\frac{\hat{\epsilon}_o}{\gamma^4\epsilon_1}+
\frac{\epsilon_1^2}{\gamma^4\hat{\epsilon}_o^2}\Bigg)
\eqno(23a)
$$
and
$$
\frac{dN_{1is~R}}{dt_1d\epsilon_1}=\frac{\pi n_o r_o^2 
c}{4\beta^6\gamma^2\hat{\epsilon}_o}\Bigg((1+\beta)\Big[\beta(\beta^2+3)+
\frac{1}{\gamma^2}(9-4\beta^2)\Big]+
$$
$$
+\frac{\epsilon_1}{\hat{\epsilon}_o}(1-
\beta)\Big[\beta(\beta^2+3)-\frac{1}{\gamma^2}(9-4\beta^2)\Big]+
$$
$$
-\frac{2}{\gamma^2}(3-\beta^2)\Big(1+\frac{\epsilon_1}{\hat{\epsilon}_o}\Big)
\ln\Big[\frac{\hat{\epsilon}_o(1+\beta)}{\epsilon_1(1-\beta)}\Big]+
\frac{\hat{\epsilon}_o}{\gamma^4\epsilon_1}-
\frac{\epsilon_1^2}{\gamma^4\hat{\epsilon}_o^2}\Bigg)~~.
\eqno(23b)
$$
We notice that the two expressions exhibit some kind of symmetry, in 
particular the 
second one can be obtained from the first one simply by reversing the $\beta$ 
sign and multiplying the whole formula by (-1). The non relativistic limit 
$\beta\rightarrow 0$ in eq.(23a,b) leads to a monochromatic spectrum, 
$\it{i.e.}$ 
a Dirac $\delta$ function, as it should be expected. Now we consider the 
ultrarelativistic limit $\beta\rightarrow 1$ and, neglecting terms smaller 
than $1/\gamma^2$ we get
$$
\frac{dN_{1is~L}}{dt_1d\epsilon_1}\approx\frac{\pi n_o r_o^2 
c}{2\gamma^4\hat{\epsilon}_o\beta^6}\Bigg[\Big(4\gamma^2+2\Big)
\frac{\epsilon_1}{\hat{\epsilon}_o}
+1-\frac{3}{2\gamma^2}-2\Big(1+\frac{\epsilon_1}{\hat{\epsilon}_o}\Big)
\ln\Big(\frac{4\gamma^2\epsilon_1}{\hat{\epsilon}_o}\Big)-
\frac{\hat{\epsilon}_o}{2\gamma^2\epsilon_1}\Bigg]
\eqno(24a)
$$
for $\frac{1}{4\gamma^2}<\frac{\epsilon_1}{\epsilon_o}\leq 1$
and
$$
\frac{dN_{1is~R}}{dt_1 d\epsilon_1}\approx\frac{\pi n_o r_o^2 
c}{2\gamma^4\hat{\epsilon}_o\beta^6}\Bigg[4\gamma^2+2+\Big(1-
\frac{3}{2\gamma^2}\Big)
\frac{\epsilon_1}{\hat{\epsilon}_o}-2\Big(1+\frac{\epsilon_1}{\hat{\epsilon}_o}
\Big)\ln\Big(\frac{4\gamma^2\hat{\epsilon}_o}{\epsilon_1}\Big)-
\frac{\epsilon_1^2}{2\gamma^2\hat{\epsilon}_o^2}\Bigg]~~~.
\eqno(24b)
$$
for $1\leq\frac{\epsilon_1}{\epsilon_o}<4\gamma^2$.
These equations must be compared with eq.(40) and eq.(44) of ref.[2]. 
We note that we have a slight difference in a couple of terms and 
in the coefficient in front of the logaritmic term. From inspection of fig.7-8 
one sees that Jones' approximated formulae exhibit a slight departure from our 
exact expressions and the difference is of order $1/\gamma^2$. Moreover it is 
also possible to notice that our eq.24 reppresent a better approximation for 
most of the $\epsilon_1/\hat{\epsilon}_o$ allowed values. At ultrarelativistic 
regime all the three kinds of expressions (eq.23, eq.24, Jones' 
approximations) are overlapping. For most applications where both non 
relativistic and ultrarelativistic regimes are of interest we consider our 
exact formula (23) which is more convenient to handle than the Jones' 
significantly more complicated expression eq.(35) in ref.[2].

\section*{\bf Conclusions}

We derived exact analytical formulae able to 
describe the ICS onto monochromatic beam lasers either in relativistic and non 
relativistic limits. Our results correct and 
extend previous known ones. These new formulae are to prefer because of 
their straightforward derivation and their more transparent form. As we have 
already shown in a separate paper [11] our results on ICS onto BBR successfully 
fit the known experimental data taken at LEP I in recent 
years. This process, ICS onto BBR, plays an important role also in 
astrophysical problems where the cosmic rays energy loss by ICS and its
consequent ICS radiation in gamma rays is of primary relevance. ICS onto 
stellar BBR may also be important in solving the well known GRB puzzle. We 
also note that the coherent nature of ICS process may significantly amplify 
the ICS itself offering a powerful diagnostic tool for the distribution of 
charges in the bunch. The optimal experimental setup should be realized when 
relativistic bunches of charges are hit by collinear back photon lasers at 
optical or infrared wavelenghts.

\section*{\bf Acknowledgements}

We wish to thank Dott. L.Silvestrini for useful comments.

\section*{\bf References}

~~\newline
[1]~V.L.Ginzburg, S.I.Syrovatskii, Soviet Phys.JETP 19 (1964) 1255.\newline
[2]~F.C.Jones, Phys.Rev. 167 (1968) 1159.\newline
[3]~G.R.Blumenthal, R.J.Gould, Rev.Mod.Phys. 42 (1970) 237.\newline
[4]~M.S.Longair, High Energy Astrophysics (Cambridge University Press, 1981).
\newline
[5]~V.S.Berezinskii, S.V.Bulanov, V.A.Dogiel, V.L.Ginzburg, V.S.
Ptuskin,\newline 
Astrophysics of Cosmic Rays (North Holland, 1990).\newline
[6]~B.Dehning, A.C.Melissinos, P.Ferrone, C.Rizzo, G.von Holtey,\newline
Phys.Lett.B 249 (1990) 145.\newline
[7]~C.Bini, G.De Zorzi, G.Diambrini-Palazzi, G.Di Cosimo, A.Di Domenico, 
P.Gauzzi, D.Zanello, Phys.Lett.B 262 (1991) 135.\newline
[8]~A.Di Domenico, Particle Accelerators 39 (1992) 137.\newline
[9]~D.Fargion, A.Salis, Nucl.Phys.B (Proc.Suppl.) 43 (1995) 269.\newline
[10]~E.Segre', Nuclei e Particelle (Zanichelli, 1994) p.137.\newline
[11]~D.Fargion, A.Salis, Preprint INFN n.1134 23/02/96, submitted to New 
Astronomy
\vspace{3.0cm}

\newpage

\centerline{\bf{Figure Captions}}

\begin{itemize}

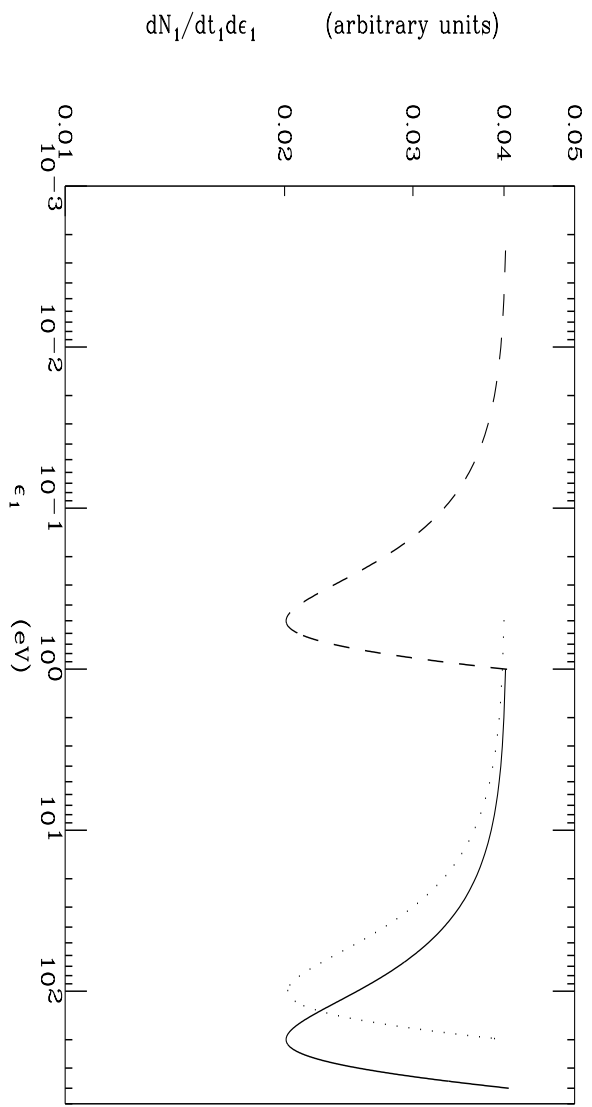
\item{\bf Fig.1}: The energy spectrum (eq.10) for $\hat{\theta}_o=0$ (dash), 
$\hat{\theta}_o=\pi/2$ (dot), $\hat{\theta}_o=\pi$ (continuous)

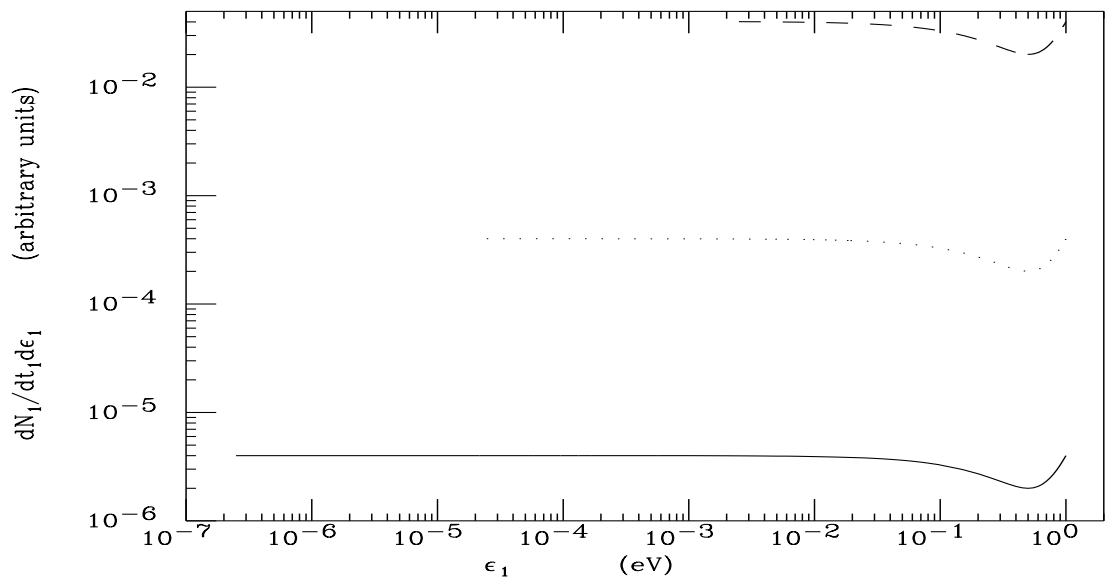
\item{\bf Fig.2}: The energy spectrum (eq.10) for $\hat{\theta}_o=0$ and 
$\gamma=10$ (dash), $\gamma=100$ (dot), $\gamma=1000$ (continuous)

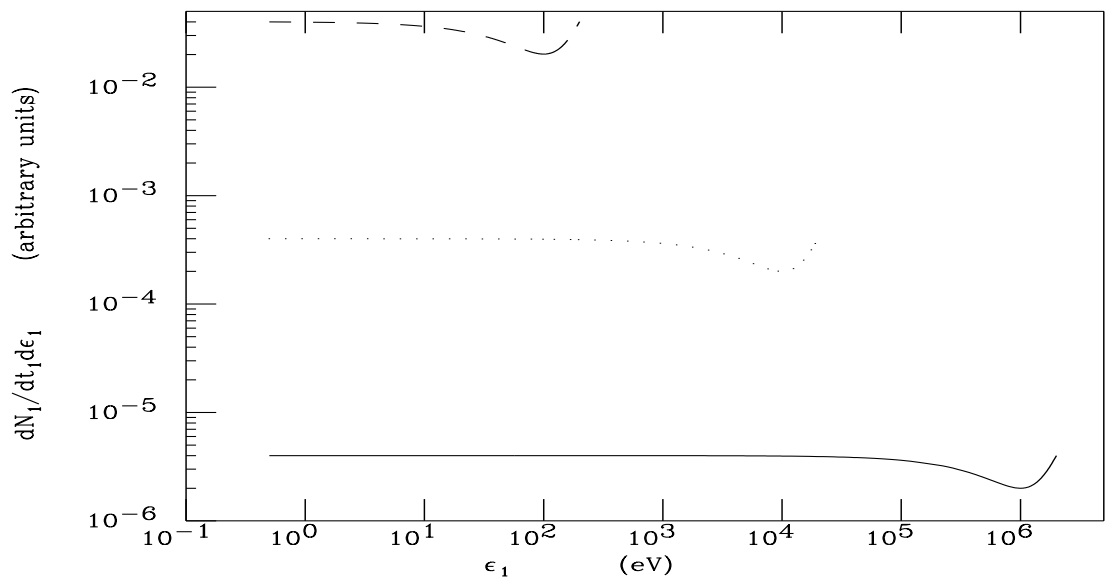
\item{\bf Fig.3}: The energy spectrum (eq.10) for $\hat{\theta}_o=\pi/2$ and 
$\gamma=10$ (dash), $\gamma=100$ (dot), $\gamma=1000$ (continuous)

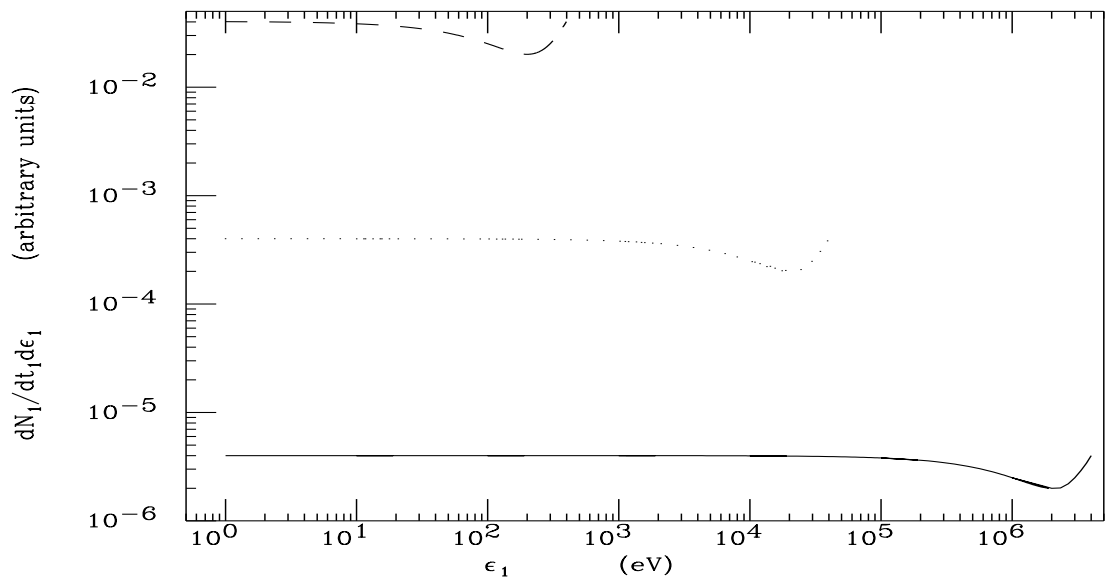
\item{\bf Fig.4}: The energy spectrum (eq.10) for $\hat{\theta}_o=\pi$ and 
$\gamma=10$ (dash), $\gamma=100$ (dot), $\gamma=1000$ (continuous)

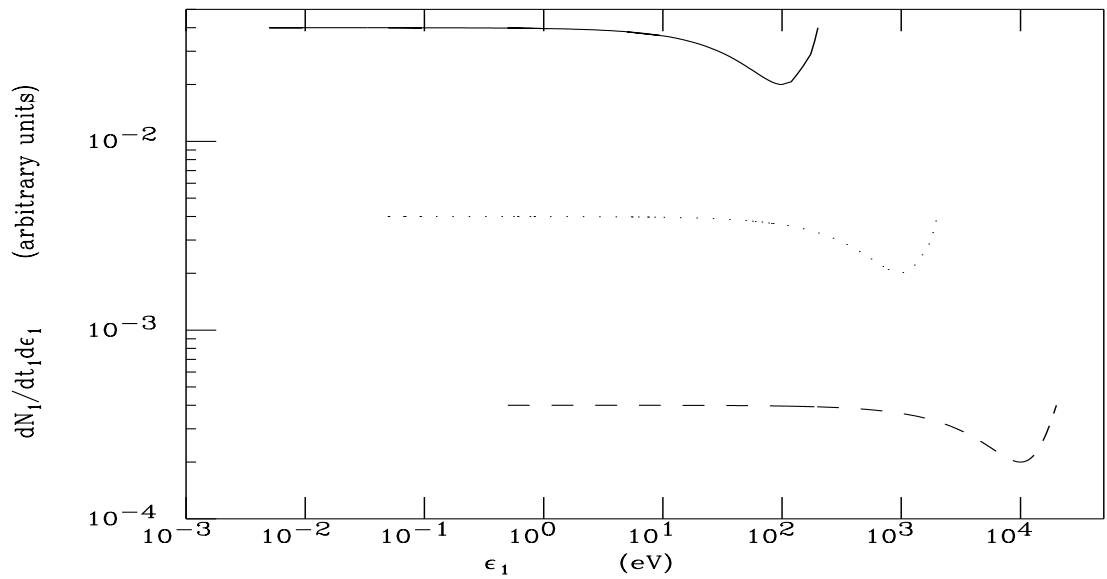
\item{\bf Fig.5}: The energy spectrum (eq.10) for $\hat{\theta}_o=\pi/2$, 
$\gamma=100$ and $\hat{\epsilon}_o=1~eV$ (dash), 
$\hat{\epsilon}_o=0.1~eV$ (dot), $\hat{\epsilon}_o=0.01~eV$ (continuous)

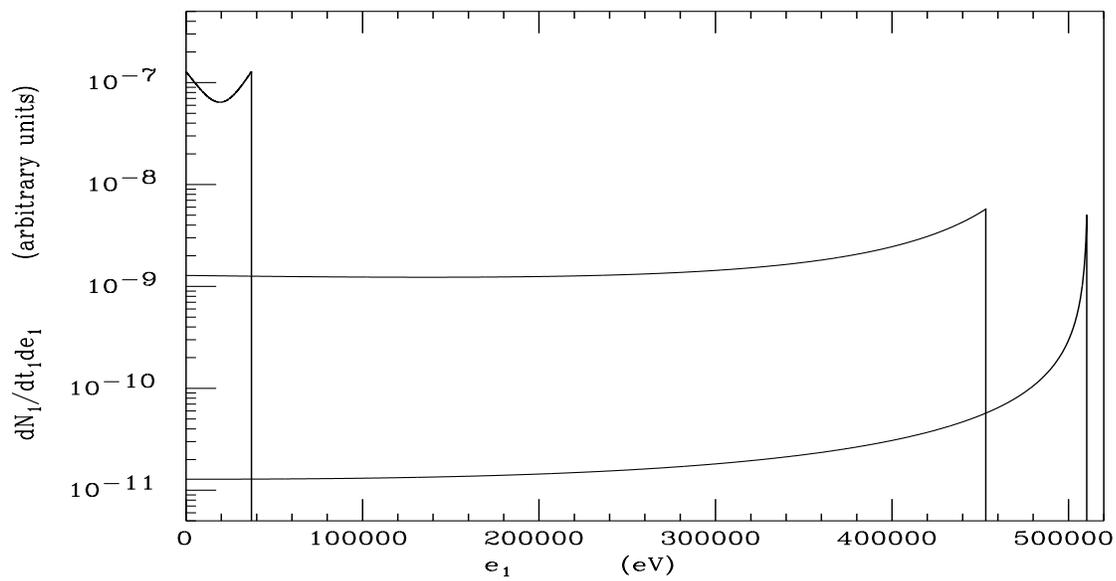
\item{\bf Fig.6}: The energy spectrum (eq.20) for $\hat{\theta}_o=\pi$, 
$\gamma=10^6$ and $\hat{\epsilon}_o=10^{-8} MeV$ (parabolic curve), 
$\hat{\epsilon}_o=10^{-6} MeV$ (asymmetric parabolic curve), 
$\hat{\epsilon}_o=10^{-4} MeV$ (peaked curve)

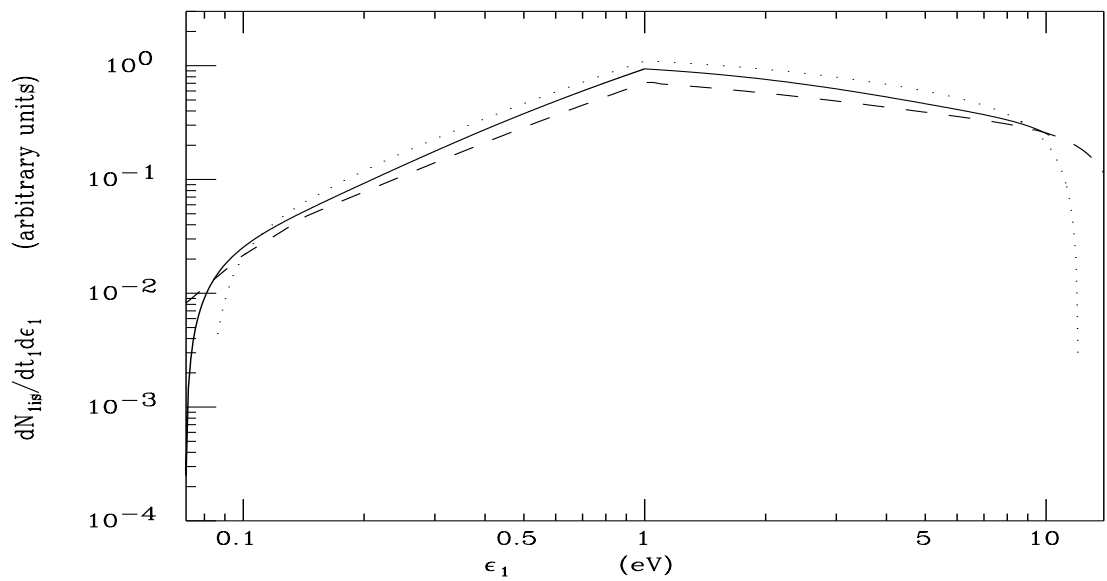
\item{\bf Fig.7}: The exact spectrum (eq.23) (continuous) and the 
approximations: Jones' (eq.40-44) (dash), eq.(24) (dot) for $\gamma=2$

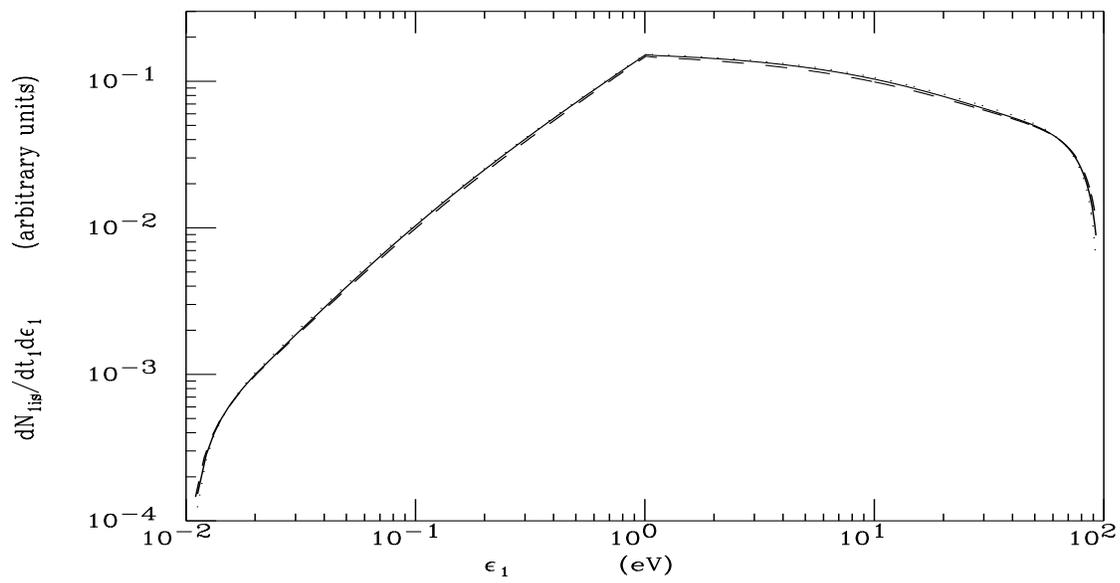
\item{\bf Fig.8}: The exact spectrum (eq.23) (continuous) and the 
approximations: Jones' (eq.40-44) (dash), eq.(24) (dot) for $\gamma=5$ 

\end{itemize}
\newpage

\end{document}